# A New Paradigm for Water Level Regulation using Three Pond Model with Fuzzy Inference System for Run of River Hydropower Plant

Ahmad Saeed[1], Ebrahim Shahzad[1], Laeeq Aslam[1,3] Ijaz Mansoor Qureshi[2,3], Adnan Umar Khan[1], Muhammad Iqbal[1,4].

[1]Department of Electrical Engineering, International Islamic University, Islamabad.
[2]Department of Electrical Engineering, Air University, Islamabad.
[3]Department of Electrical Engineering, ISRA University, Islamabad.
[4]KIOS Research and Innovation Center of Excellence and the Department of Electrical and Computer Engineering, University of Cyprus, Cyprus, Nicosia.

**Correspondence**
Laeeq Aslam, Department of Electrical Engineering, ISRA University, Islamabad
Email: laeeq.msee424@iiu.edu.pk

**Summary:** *The energy generation of a run of river hydropower plant depends upon the flow of river and the variations in the water flow makes the energy production unreliable. This problem is usually solved by constructing a small pond in front of the run of river hydropower plant. However, changes in water level of conventional single pond model results in sags, surges and unpredictable power fluctuations. This work proposes three pond model instead of traditional single pond model. The volume of water in three ponds is volumetrically equivalent to the traditional single pond but it reduces the dependency of the run of river power plant on the flow of river. Moreover, three pond model absorbs the water surges and disturbances more efficiently. The three pond system, modeled as non-linear hydraulic three tank system, is being applied with fuzzy inference system and standard PID based methods for smooth and efficient level regulation. The results of fuzzy inference system are across-the-board improved in terms of regulation and disturbances handling as compared to conventional PID controller.*

## List of Symbols and Abbreviations:

The symbols used are as follows,
$H_i$ = Height of pond in meters, where $i = 0, 1, 2$
$x_i$ = Water level of pond in meters, where $i = 0, 1, 2$
$H_s$ = Height of Surge Tank
$X_s$ = Level of water in Surge Tank
$n$ = Dimensionless Water loss coefficient = 0.98
$g$ = gravitational constant.
$a$ = Cross sectional area of ducts between $T_1$ and $T_0$ and $T_2$ and $T_0$
$A$ = Area of the Ponds $T_0$, $T_1$ and $T_2$ are same
$A_s$ = Area of Surge Tank.
$Q_t$ = Water flow rate in the headrace.
$L_t$ = Length of headrace.
$C_t$ = Friction constant of headrace. = 0.98
$A_t$ = Cross sectional area of head race.

## 1. INTRODUCTION

Rivers and the water bodies contains kinetic and potential energy. This energy is transformed to mechanical energy and then to electricity which is known as hydroelectric power or hydropower in popular terms. Hydropower has been in use since ancient times by use of water wheels to provide power to grind mills and other industrial tools. It provides almost 20% of the world's energy requirements in the modern world [1] which is considered to be the most cost effective source of energy, comprising nearly 80% of all the renewable energy. Hydro power plants (HPPs) can be classified as conventional and non-conventional HPPs where conventional power plants consist of a dam, lake, penstock and a power house; however; non-conventional ones include run of river power plants, tidal and offshore wave power plants. This classification is based on size, water head, storage capacity and type of generation facility. Today majority of the HPPs in practice are conventional HPPs. However, since the suitable sites for conventional HPPs are limited and have far-reaching impacts on eco-system; therefore, small and non-conventional methods for harnessing hydro-energy are being utilized now a days with the increasing research trends of distributed power based smart grids. One of the frequently used non-conventional HPPs are Run of River Hydro power plant (ROR HPP).

In run of ROR HPPs the reservoir is mostly absent or alternatively there is a very small pond or reservoir; however one of the major issues with ROR HPPs is its inconsistent output due to weather and temperature changes resulting in changing river's rate of flow [2]. The available power may be at peak during the rainy season whereas in the dry season the available power may be less. The water level is regulated by constructing a weir at the riverbed. In some sites where river water is redirected to flow in power plant, requires construction of a small pond at the head of plant to provide it with some energy storage capacity. So it is different from conventional power plants in terms of very lesser power storage capacity.

The existing literature on ROR HPPs can be categorized in three different categories i.e. control of plants, power forecasting and environmental economics of the power plant. Control of ROR HPP is done using classical control [3] whereas head pond level control is done using fuzzy inference system [4]. Automatic generation control of ROR HPP can be developed in two modes; constant power mode and constant frequency mode [5], which is a more generic control methodology. Control of a ROR HPP with variable speed Kaplan turbine [6] has been devised and experimentally tested. A novel idea to complement the generated power with a bank of super capacitors is proposed recently it devises a methodology to effectively integrate the capacitor bank to the ROR HPP [7]. A software based troubleshooting / fault diagnostic tool [8] is developed for usage in run of river hydropower plant which is truly not a fault diagnostic system, instead it is an end-user based tool which simply indicates the possible reasons for out of ordinary system behavior. A real river flow data is used to simulate a run of river hydropower plant [9], while another work develops an electromechanical system for regulating generation frequency in the presence of electronic load controllers [10].

The work in [10,11] uses a mechanical gearbox speed increaser to increase the speed of a turbine for higher efficiency[11] along with regulation of output voltage using AC-DC-AC converters. The works in [12–17] propose different ways to integrate run of river hydropower plants to the grid and with other power plants such as wind power. They also optimize the power generation such that the losses are minimized, while [18] proposes establishment of independent micro-grids near run of river power plants.

Scheduling the ROR hydropower production due to varying power demands is also a very trending topic in this area of research. Different types of power production scheduling is discussed in [19–22] ranging from using reversible pumps [19] to creating a power production function [20] based short term scheduling using historical weather data. Forecasting accuracy is discussed in [23] and [24] which uses Artificial Neural Network (ANN) to decide factors to be included in forecast related decision making. Another work in [25] also analyzes the inclusion of water continuity equations for forecasting while controlling the water level in the power plant, where the author concludes that fluid continuity equations as unnecessary for short term forecast.

The works in [26–28] have also discussed the economic and environmental impacts of ROR HPPs. Cost economics optimization [29] and predicting the cost overrun [25] has also been done for run of river hydropower plants, making the run of river hydropower more sustainable and cost effective [17,30–32].

The main contributions of the paper are summarized as follows.

- A dynamic model of the three pond ROR HPP system is proposed instead of the traditional single pond system, which is basically inspired from three tank model. It makes it more robust against seasonal variations of water supply and gives the plant more flexibility against the fluctuating load demands which is shown in simulation results.
- The levels of the ponds are controlled using fuzzy inference engine (FIS). The control technique aims to provide a redirected ROR HPP with faster control (regulation) as compared to traditional ROR HPPs.

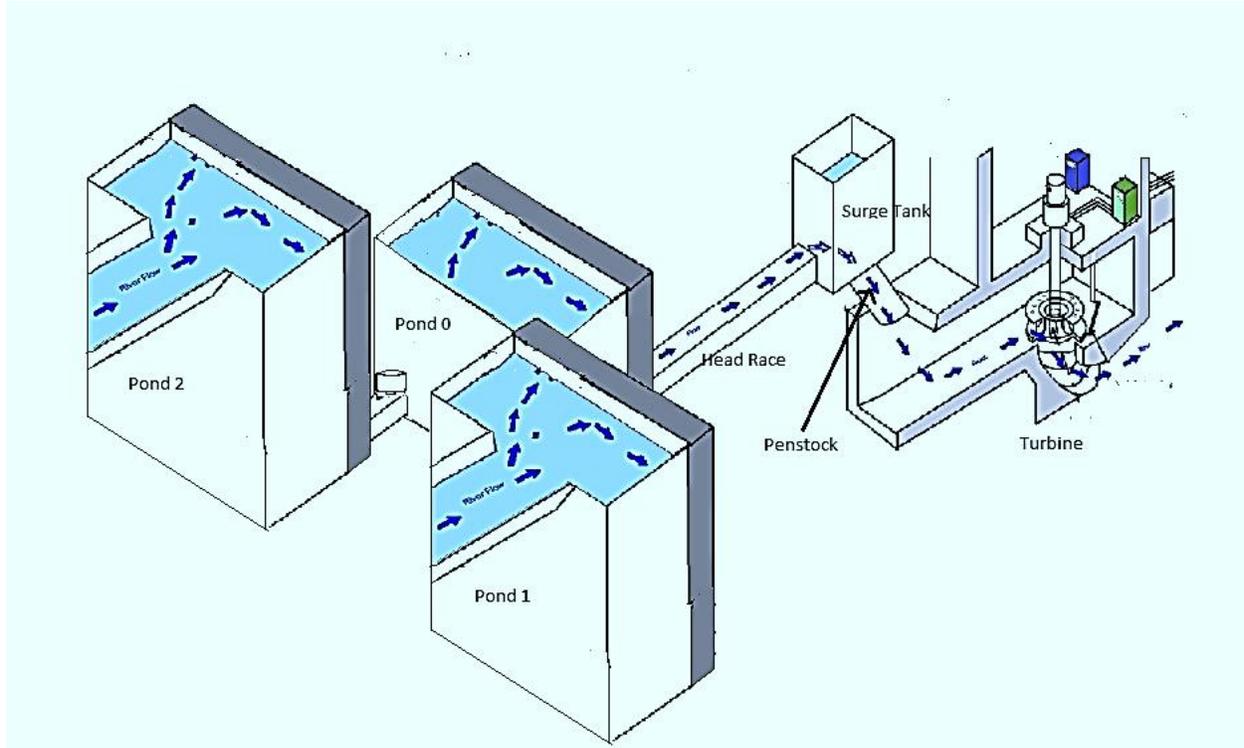

Figure 1 Proposed Three Pond run of river hydropower plant.

## 2. SYSTEM MODEL

In a traditional redirected ROR HPP, a weir or a small wall is constructed on the riverbed to regulate the water level. Weir is different from a dam in such a way that water is meant to flow over the weir. A channel is made in the river and the water is flown towards a pond. This pond is the head pond of the system and is connected to the penstock or headrace (in the presence of a surge tank).

In our run of river hydropower plant (Figure 1), there are three head ponds or tanks $T_0$, $T_1$ and $T_2$. Water from the river is allowed to flow in the ponds $T_1$ and $T_2$. The ponds $T_1$ and $T_2$ are connected at the bottom to pond $T_0$,. The pond $T_0$, is connected to the head race tunnel which goes to penstock and turbine via surge tank. The connecting tunnels between $T_1$, $T_0$, and $T_2$, $T_0$, have controllable valves $s_1$ and $s_2$. Also note that Figure 1 is originally taken from [1,2] being a single reservoir model and modified for our requirement as a three pond system.

The system is modeled as a set of equations while dealing with each part of the system. As our hydro-system consists of three ponds, a head race and a surge tank, we have a set of five equations. Structurally, ponds and tank have the same mathematical model where the rate of change of water level is a function of the total flow in and out of the pond or tank. As the headrace is a rather long tunnel, the flow depends upon the water levels at the either end of the headrace. The system equations are given below.

The equation for Pond 0 is

$$\frac{d}{dt}x_0 = \frac{n_{\!a}}{A}\left[s_1\sqrt{2g|x_1 - x_0|}sgn(x_1 - x_0) + s_2\sqrt{2g|x_2 - x_0|}sgn(x_2 - x_0)\right] - \frac{Q_t}{A} \quad (1)$$

where $sgn(a)$ is defined as:

$$sgn(a) = \begin{cases} 1 & a > 0 \\ -1 & a < 0 \end{cases}$$

and $|a|$ is defined as:

$$|a| = \begin{cases} a & a > 0 \\ -a & a < 0 \end{cases}$$

The equation for Pond 1 is

$$\frac{d}{dt}x_1 = -\frac{n_{\!a}}{A}s_1\sqrt{2g|x_1 - x_0|}sgn(x_1 - x_0) + \frac{U_1}{A} \quad (2)$$

The equation for Pond 2 is

$$\frac{d}{dt}x_2 = -\frac{n_{\!a}}{A}s_2\sqrt{2g|x_2 - x_0|}sgn(x_2 - x_0) + \frac{U_2}{A} \quad (3)$$

The equation for headrace

$$\frac{d}{dt}Q_t = \frac{gA_t}{L_t}(x_0 - x_s) - C_tQ_t|Q_t| \quad (4)$$

The surge tank equation is

$$\frac{d}{dt}x_s = \frac{Q_t}{A_s} - \frac{n_{\!a}}{A_s}\sqrt{2gx_s} \quad (5)$$

These five equations can be combined to form a nonlinear system model as

$$\frac{d}{dx}\vec{x} = \vec{f}(x, u) \quad (6)$$

Where the states variables are

$$\vec{x} = [x_0 \; x_1 \; x_2 \; Q_t \; x_s]^T \quad (7)$$

And the control variables are

$$\vec{u} = [U_1, U_2, s_1, s_2]^T \quad (8)$$

The outputs are $x_1$, $x_2$. The output equation is

$$y = Cx \quad (9)$$

Where C is defined as

$$C = \begin{bmatrix} 1 & 0 & 0 & 0 & 0 \\ 0 & 1 & 0 & 0 & 0 \\ 0 & 0 & 1 & 0 & 0 \\ 0 & 0 & 0 & 0 & 0 \\ 0 & 0 & 0 & 0 & 0 \end{bmatrix} \quad (10)$$

where $U_1$ and $U_2$ are the controllable water flow rates into the ponds $T_1$ and $T_2$ and $s_1$ and $s_2$ are the valves between the ponds $P_1$ and $P_0$ and $P_2$ and $P_0$ respectively. $U_1$ and $U_2$ can be varied between 0 to $U_{max}$, while the valves $s_1$ and $s_2$ are varied from 0 to 1. The system structure in simulink is shown in Figure 2.

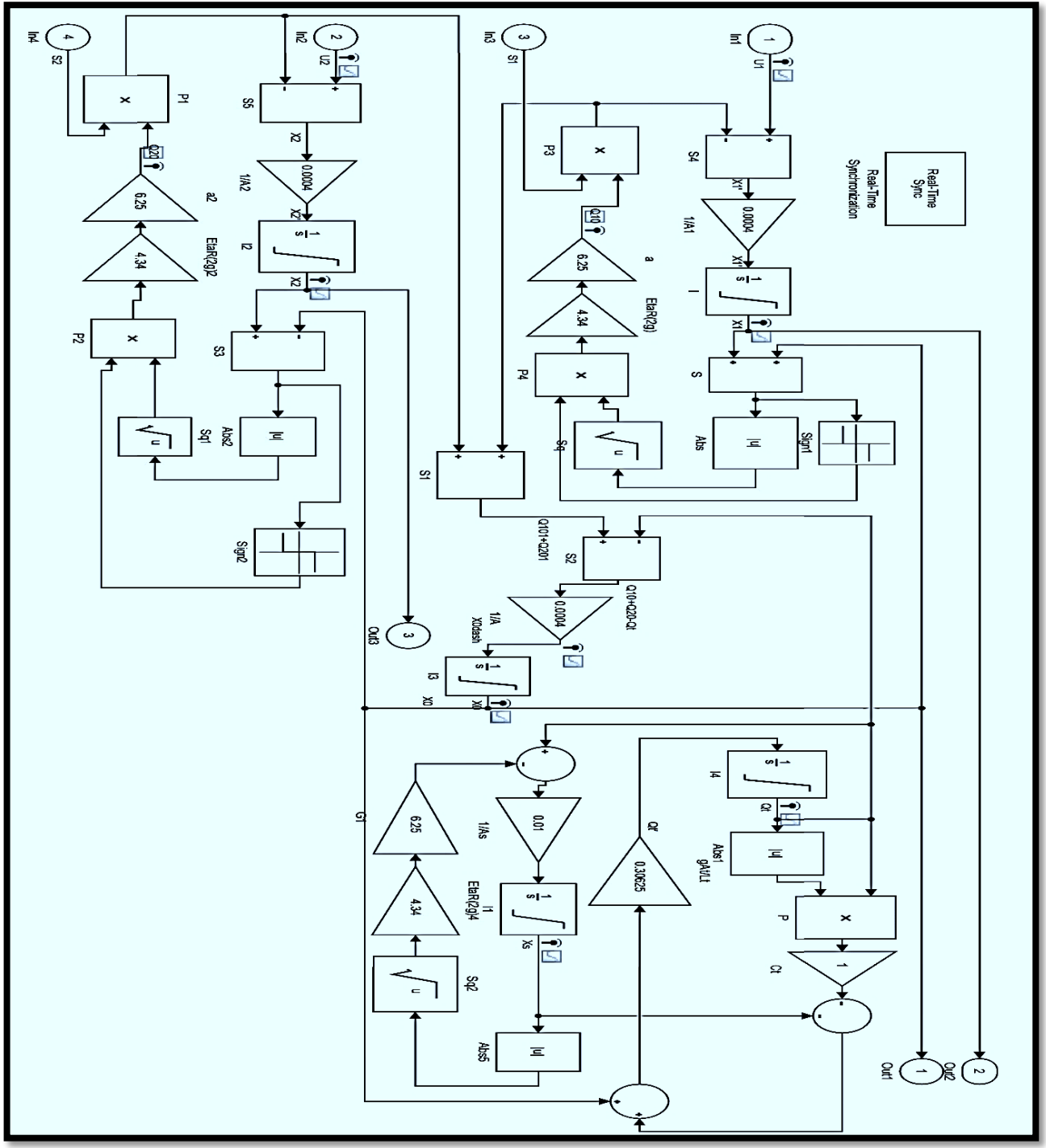

**Figure 2:** System model in Simulink

## 3. CONTROL DESIGN

The objective is to maintain water levels in the ponds to a specific predefined height. In this work system is different from traditional ROR system, since it proposes three ponds instead of using a single pond, so this work is only interested in the water dynamics of the system without the inclusion of turbine. The proposed three pond system is better than traditional single pond system for maintaining the water level in the head ponds using Fuzzy Inference Engine (FIS), while PI control is also applied on both cases (traditional and proposed) for the sake of comparison. So primarily fuzzy controller is used and then the response is compared with PID controller.

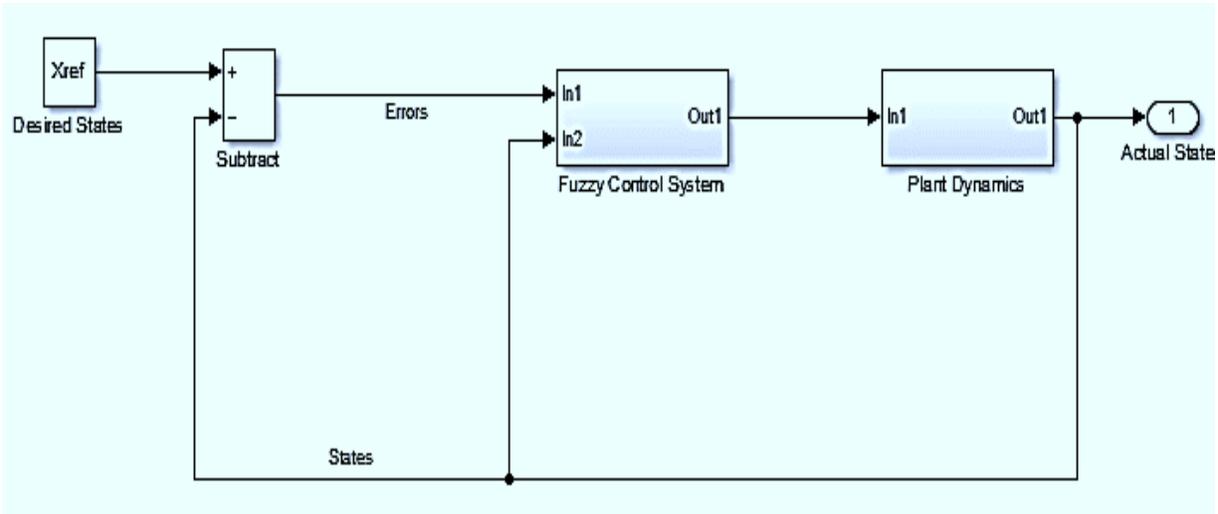

Figure 3: Control scheme of system

### 3.1 Fuzzy Control

The control scheme used is like a traditional control system with fuzzy controller. The only difference between our scheme and traditional scheme is that the fuzzy controller is driven by both the errors and output states of the system. In Figure 3 the simplified control strategy is shown.

In the previous section, it is described the system with our control variables $U_1, U_2$ $s_1$ and $s_2$, while out of the five states it is required for $x_0$, $x_1$ and $x_2$ to track the desired levels respectively. The control system consists of two parts, a set of three subtractors, and a set of two fuzzy inference engines. The detailed control blocks in Simulink are shown in Figure 4.

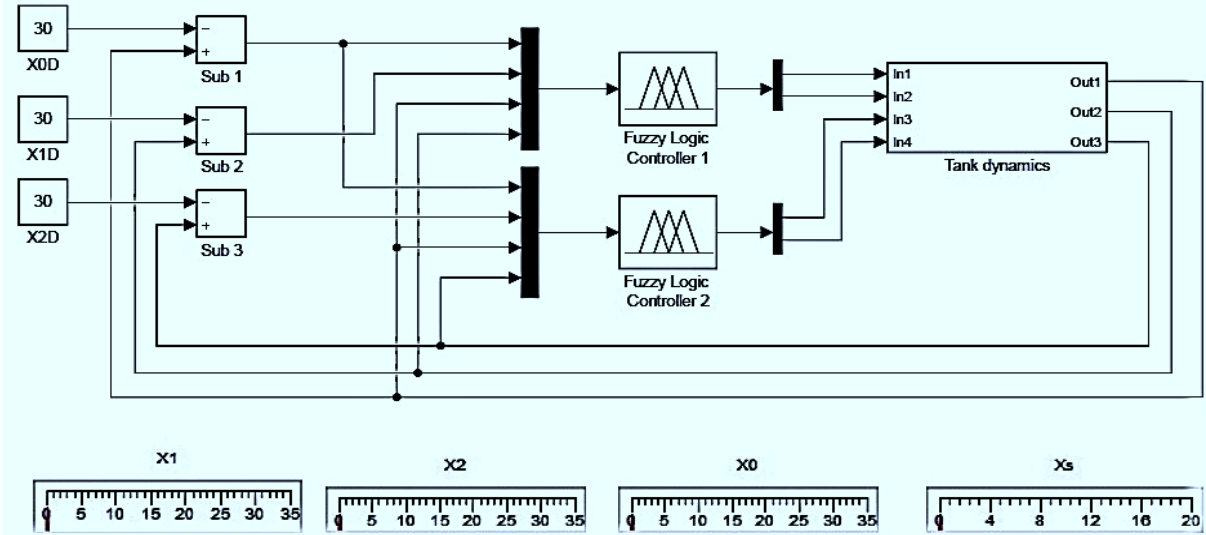

Figure 4 Fuzzy Control Scheme in Simulink

The subtractors simply determine the difference between the present states and the desired state values. The advantage of using subtractors is the ability to change desired levels at run time without needing a new set of fuzzy rules for the fuzzy inference engine. The outputs of the subtractors group are $E_0$, $E_1$ and $E_2$ errors of the ponds. Depending on the errors being negative, zero or positive, the fuzzy inference engine is able to make decisions.

The second and the most important part of the control system is the actual fuzzy inference system (FIS). A fuzzy inference system consists of set of IF-THEN rules. For example, for a two input one output system the rules are of the form:

$$Ru^1: IF\ x_1\ is\ A_1^1\ and\ x_2\ is\ A_2^1\ THEN\ y\ is\ B^1 \tag{11}$$

For example if a two input system has 4 memberships for both inputs then a complete set of rules for the same system will consist of $4 \times 4$ or $4^2$ rules. The number of output memberships can be the same as the number of rules or less.

For the considered pond system in this paper we require to regulate the water level in the ponds to desired value, so the fuzzy inference engine is given the actual water levels and error values. Based on the actual and error values our FIS either increases the flow of water $U_1$ and $U_2$ in case the error is negative, maintains the inflow and outflow rate in case the errors are zero and cuts off the inflow while encourages the outflow in case error is positive. FIS is fed with three errors $E_0, E_1, E_2$ and three levels $x_0, x_1, x_2$. To make the system simpler, the heights are assigned membership functions *LOW, MEDIUM* and *HIGH*. While membership functions assigned to errors are *NEGATIVE, ZERO* and *POSITIVE*. All of these membership functions are self-explanatory and the ZERO membership function actually mean near zero error. In order to have some degree of precision, each output has five levels from 0 to 4. '0' is stop while '4' is full throttle. As stated above, for the rule base to be complete we have to devise $3^6$ or 729 rules to run the system effectively. While defining this system in the Matlab fuzzy toolbox, each rule consists of six inputs and four outputs and each input and output has three and five levels respectively. This translates into a huge mathematical overhead that makes the simulation too slow to be simulated. Therefore to avoid this huge overhead we replace our single fuzzy system with two parallel fuzzy systems. Fuzzy System 1 ($FS_1$) deals with Pond 0 and Pond 1 while regulating $U_1$ and $s_1$, whereas Fuzzy System 2 ($FS_2$) deals with Pond 0 and Pond 2 and regulates $U_2$ and $s_2$. So effectively it has four inputs and two outputs for each fuzzy system. So, the number of rules required also drops to $3^4$ or 81 for each system. So we have two systems with 81 rules each.

First it is needed to look at the rules of fuzzy system 1, justify its basis and make a full set of 81 rules for it and then use the same rules for fuzzy system 2, as both of the systems are essentially the same with the exception of $x_1$. The fuzzy system has four inputs, two errors which can be negative zero or positive and three levels which can be low, medium or high. The objective of the fuzzy system is to push the error levels to zero while controlling the outputs $U_1$, which is the water flow into pond 1 and $s_1$ which is the valve between ponds 1 and 0. Starting from zero initial conditions, both $x_1$ and $x_0$ are at LOW and both $E_1$ and $E_0$ are negative so the output is to have a high inflow rate $L_4$ at $U_1$ and the valve $s_1$ is also to be set at open $s_4$. If both the errors are zero and both the heights are at similar level then it is a matter of balancing the inflow with outflow, i.e. both $U_1$ and $s_1$ are at intermediate positions to keep the system in equilibrium. If there is positive error then the inflow is stopped at $L_0$ and depending upon whether the error is positive at $E_1$ or $E_2$ the valve $s_1$ is either at full open or full closed position. If the desired level is such that $X_{1d} < X_{0d}$ then after reaching zero at $E_1$ both the inflow $U_1$ and Valve $s_1$ is closed and the task for having the error $E_2$ to reach zero is left to the second fuzzy controller. Using above guidelines we generate a set of 81 fuzzy rules for both of the fuzzy inference engines $F_1$ controlling the levels of pond 1 and pond 0, while fuzzy inference engine $F_2$ controls the levels of pond 2 and pond 0; there is no conflict between $FS_1$ and $FS_2$ as the desired level of pond 0 is fed to both the fuzzy systems ($FS_1$ and $FS_2$) by the same source. Now a rule is given as example.

*IF E1 is NEGATIVE and E0 is NEGATIVE and X1 is LOW and X0 is LOW THEN U1 is L4 and S1 is V4*

This rule relates to the fuzzy system 1 as it deals with heights and errors of pond 1 and pond 0. The first part of the rule states that if $E_1$ and $E_0$ are both negative, it means that the water level is less than desired for both the ponds. The second part states that both $x_1$ and $x_0$ are low, which is self-explanatory. Now it is desired for errors to reach to zero, so we must have a large flow in both $x_1$ and $x_0$ so $U_1$ is $L_4$, as $U_1$ is fed directly to $x_1$, it is required that water flows freely into $x_0$ as well so the valve $s_1$ is at $V_4$ i.e. full throttle.

The generalized form of the output of the fuzzy controller is given by

$$u(t) = \frac{\sum_{j=1}^{81} u_j \left( \prod_i^4 \mu_{A_i^j}(k_i) \right)}{\sum_{j=1}^{81} \left( \prod_i^4 \mu_{A_i^j}(k_i) \right)} \tag{12}$$

Where $u(t)$ is the output of the controller for $U_1$ or $s_1$, $k_i$, is the input value of $x_0, x_1, E_0$ and $E_2$. A is the linguistic variable for input which can be $NEG, ZERO, POS$ or $LOW, MED, HI$. $u_A$ is the membership function of A and $u_j$ is the firing strength of $j^{th}$ fuzzy rule.

### 3.2 PID Control

The PID control scheme used is of the same layout as a traditional control scheme i.e. the control action is driven by the error of the system. This work proposes three controllers 0, 1 and 2. The output of the controllers 1 and 2 are $U_1$ and $U_2$ which regulate the water levels of pond 1 and 2 respectively. The output of controller 0 regulates the level of pond 0 and its output drives both $s_1$ and $s_2$.

The generalized form of the output of the PID controller is given by

$$u(t) = K_P e(t) + K_I \int_0^t e(t)dt + K_D \frac{d}{dt} e(t) \tag{13}$$

Here $e(t)$ is the error signal, while $K_p$, $K_I$ and $K_D$ are the proportional, integral and derivative constants respectively.

## 4. RESULTS & COMPARISONS

For analyzing the proposed system simulation is performed using Simulink (Matlab) and compared it to the standard redirected run of river hydropower plant. For the sake of comparison, this work compares three pond model with a traditional single pond model having the same capacity. The total storage capacity of the three pond system is equal to the storage capacity of traditional single pond system.

This work arbitrarily uses the following values of proposed model parameters in the simulation as a case study.

$H_0 = H_1 = H_2$ = Height of pond 0, 1, 2 = 35 m

$H_s$ = Height of Surge Tank = 15 m

$a$ = Cross sectional area of ducts between $T_1$ & $T_0$, $T_2$ & $T_0$ = $6.25 m^2$

$A$ = Area of the Ponds $T_0, T_1, T_2$ = $50m x 50m = 2500 m^2$

$A_s$ = Area of Surge Tank. = $10m x 10m = 100 m^2$

$L_t$ = Length of headrace. = $200 m$

$A_t$ = Cross section al area of head race = $6.25 m2$

$U_{max}$ = Maximum controllable inflow of water in ponds $T_1$ and $T_2$ = $100\ m^3/sec$

Traditional System: Single Pond Model

Proposed System: Three Pond Model

$T_0, T_1, T_2$: pond 1, pond 2, pond 3

For the fair comparison of three pond model with traditional single pond model, all the parameters are same except, the A matrix of traditional model is taken thrice and $U_{max}$ is double than that of three pond model for storage capacity equivalence. The levels of both the systems are controlled by fuzzy inference control.

### 4.1 RESULTS

In this section results of the proposed FIS for controlling water level in three pond model are discussed in which this work considers four different cases. In first case the desired water level in all three ponds is same. Whereas, in second case desired water level in pond 0 is less than the target water level in pond 1 and pond 2, while the desired water level of pond 1 and pond 2 is equal in this case. In the third case desired water level in all three ponds is different from each other and is chosen such that desired water level of pond 0 is less than the level of pond 1 which is less than the desired water level of pond 2. In case 4 water levels are same as in case 1 but PID controller is used instead of using FIS to achieve the desired water level. The results of these four cases are discussed as follows.

#### 4.1.1   Case I: ($T_0 = 30, T_1 = 30, T_2 = 30$, Fuzzy Controlled)

The desired water level is set at 30 meter for the proposed three pond system with zero initial conditions. It is observed that all the states take nearly half an hour to reach steady state values. The steady state error is 0.5 meter and the time taken to rise to the steady state is somewhat different for ponds $T_1$ and $T_2$ in comparison to pond $T_0$. The ponds $T_1$ and $T_2$ reach the steady state in nearly 27 minutes while pond $T_0$ gets to steady state in almost 30 minutes. These results are graphically represented in Figure 5 where increase in pond water level is plotted against time. In Figure 5 blue line represents increase in pond 0 water level, dotted red line represents increase in pond 1 water level and green line represents increase in pond 2 water level. Results show that the proposed scheme is capable of maintaining water levels on desired levels.

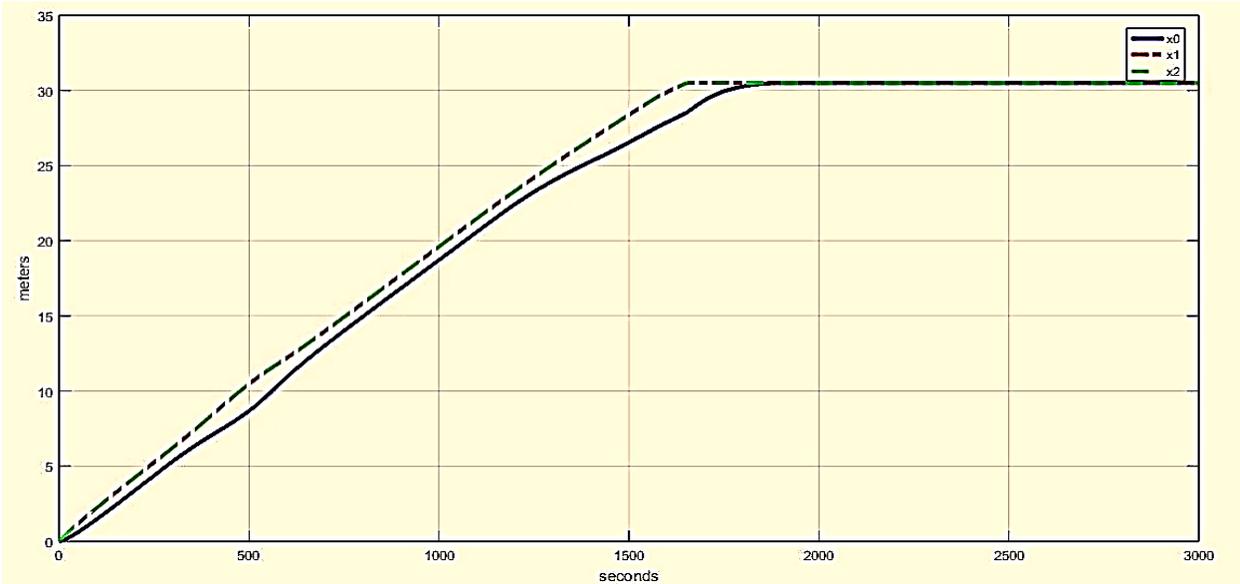

Figure 5 Desired Level vs. Time

4.1.2  Case II: (($T_0 = 25, T_1 = 30, T_2 = 30$, Fuzzy Controlled)

In this case the desired level in pond zero is 25 meter whereas for pond 1 and 2 is 30 meters. The system reaches the desired levels of 30 meters in pond 1 and pond 2 within 27 minutes. However, as the desired level for pond 0 is 25 meters, the desired level is achieved in 24 minutes the steady state error is 0.5 meter. These results are graphically shown in Figure 6 where pond water levels are plotted against time. Water level of pond 0, 1 and 2 is represented by blue, green and red dotted line respectively.

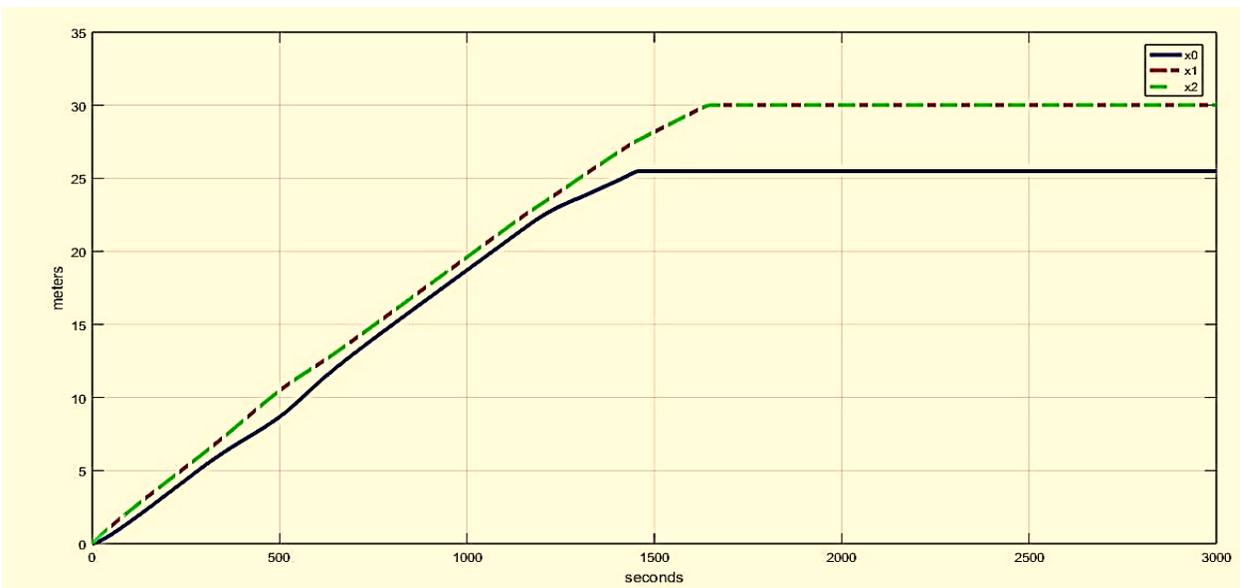

Figure 6 Desired Level vs. Time

### 4.1.3 Case III: ($T_0 = 20, T_1 = 25, T_2 = 30$, Fuzzy Controlled)

In this case the desired water level for pond 0, 1 and 2 are 20, 25 and 30 meter respectively. Overall system reach steady state in less than 30 minutes. To reaches desired level of 20 meter in nearly 18 minutes while $T_1$ and $T_2$ reach their desired level of 25 and 30 meters in less than 22 and 27 minutes respectively. Moreover only $T_0$ shows the steady state error of nearly 0.5 meters while the rest two show negligible steady state error. The results are graphically shown in Figure 7.

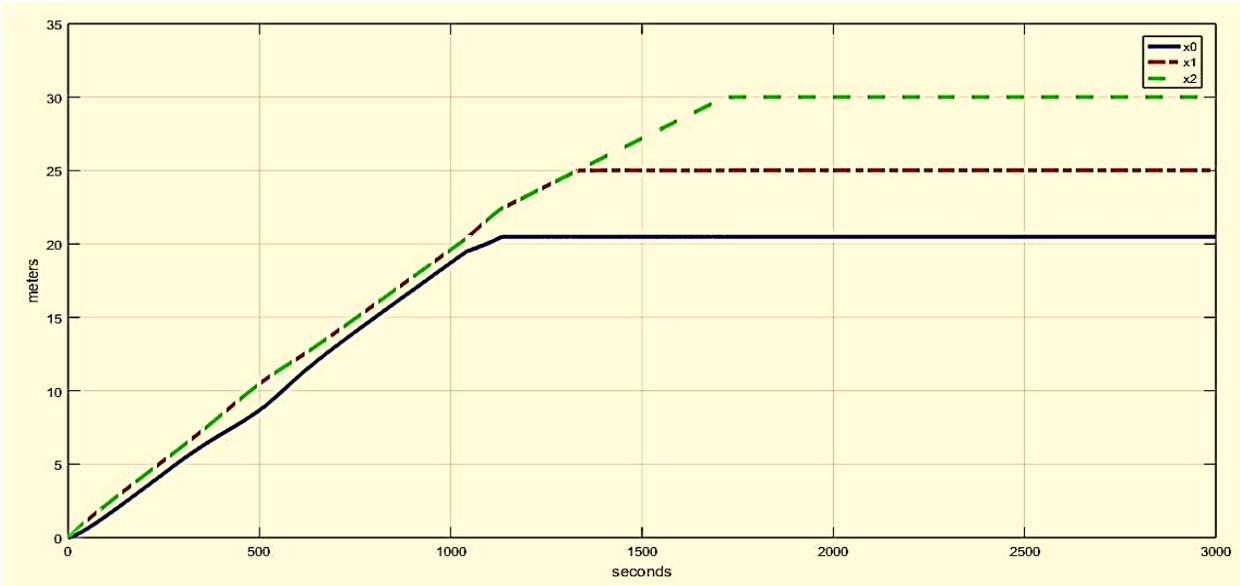

Figure 7 Desired Level vs Time

### 4.1.4 Case IV: ($T_0 = 30, T_1 = 30, T_2 = 30$, PID Controlled)

The first simulation is repeated while replacing the PID controller. With zero initial conditions the water level in the ponds reach the required height is less than half an hour but it exhibits an overshoot of nearly 4 meters and then slowly settles to the required level. Like as stated before the output of our controller varies from 0 to $U_{max}$ for $U_1$ and $U_2$ and from 0 to 1 for $s_1$ and $s_2$ therefore the simulation shows such behavior of the system. The results are graphically shown in Figure 8.

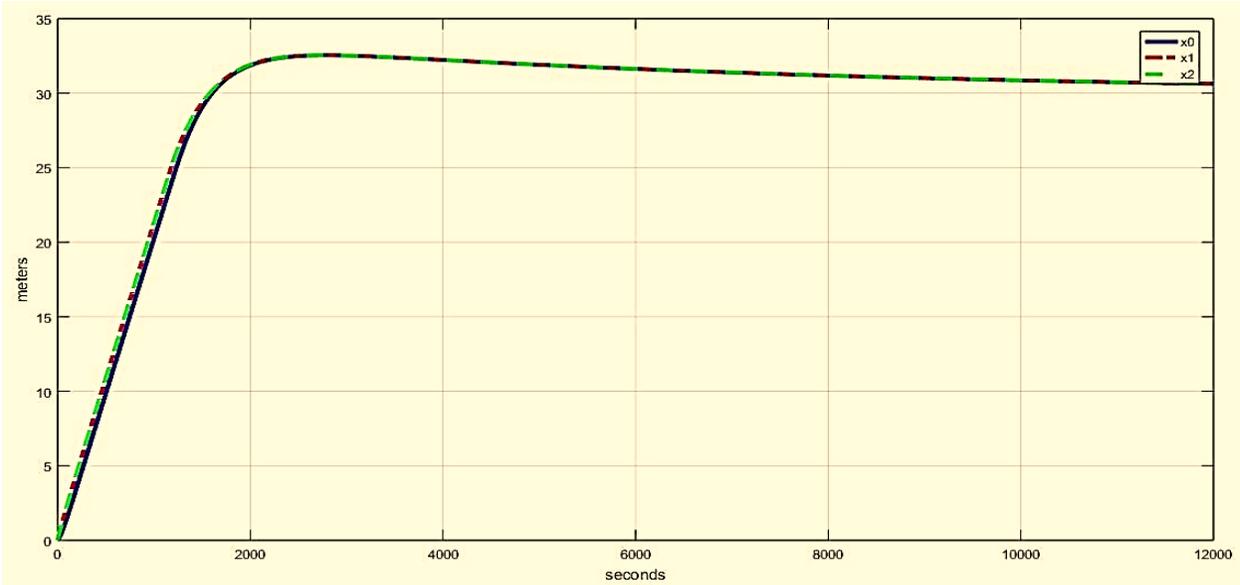

Figure 8 Desired Level vs Time

Then simulations are repeated with different heights for different ponds. It has been observer that the pond with highest water level manages to maintain its level while rest of the ponds exhibit an oscillatory behavior for 2-3 hours before settling.

### 4.2 COMPARISONS:

This section compares the proposed work with existing work in four different prospects. In case 1 proposed three pond model is compared with the traditional single pond model using a PID controller. In case 2 FIS controller is compared with the PID controller for three pond model. Case 3 compares three pond model with single pond model under the effects of sinusoidal disturbances. Finally, case 4 compares single pond and three pond model response for a flash flood case. These four comparison cases are discussed as follows.

4.2.1  Case I: Three pond system vs Traditional Single Pond Model using PID Control

While comparing the two systems in normal conditions for PID controller we see that while the three pond system is faster by a few minutes, it also exhibits a slightly larger overshoot than the traditional system. The results are graphically shown in Figure 9.

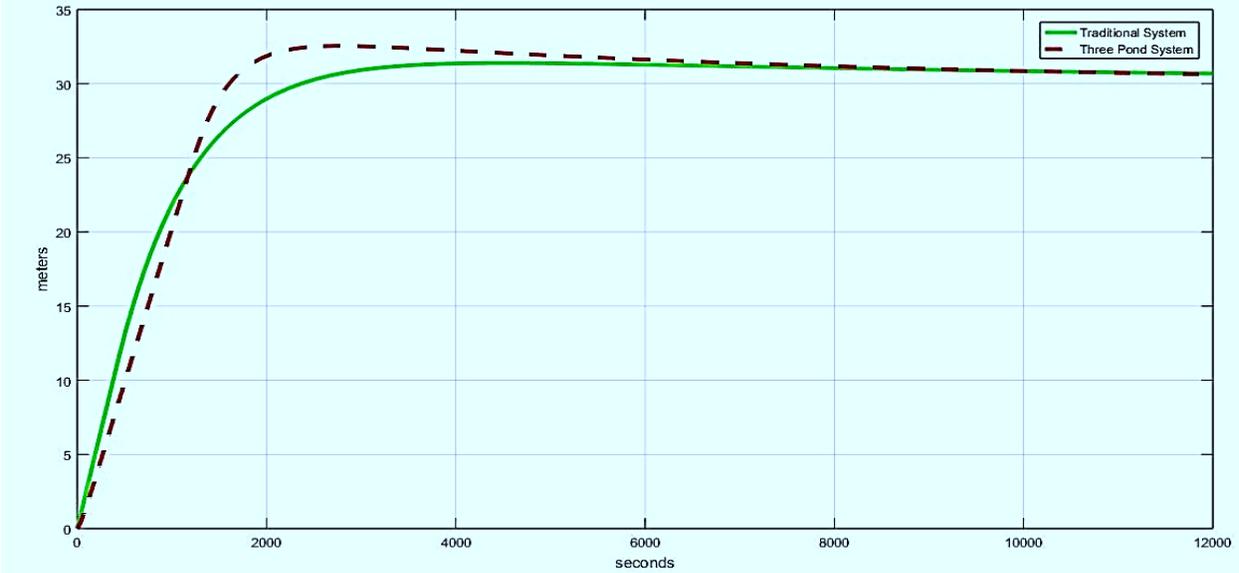
Figure 9 Comparison of proposed system with a traditional system for PID controller,

### 4.2.2 Case II: Fuzzy Controlled vs PID Controlled three pond system

While comparing the two controllers FIS vs PID for three pond system, the advantage of fuzzy controller over PID is overwhelming in every metrics. The system reaches its steady state values in nearly half an hour without having a measureable overshoot while PID controller requires 1-2 hours before settling. Therefore for comparing both the systems against disturbances we ignore PID controller in favor of fuzzy. The results are graphically shown in Figure 10

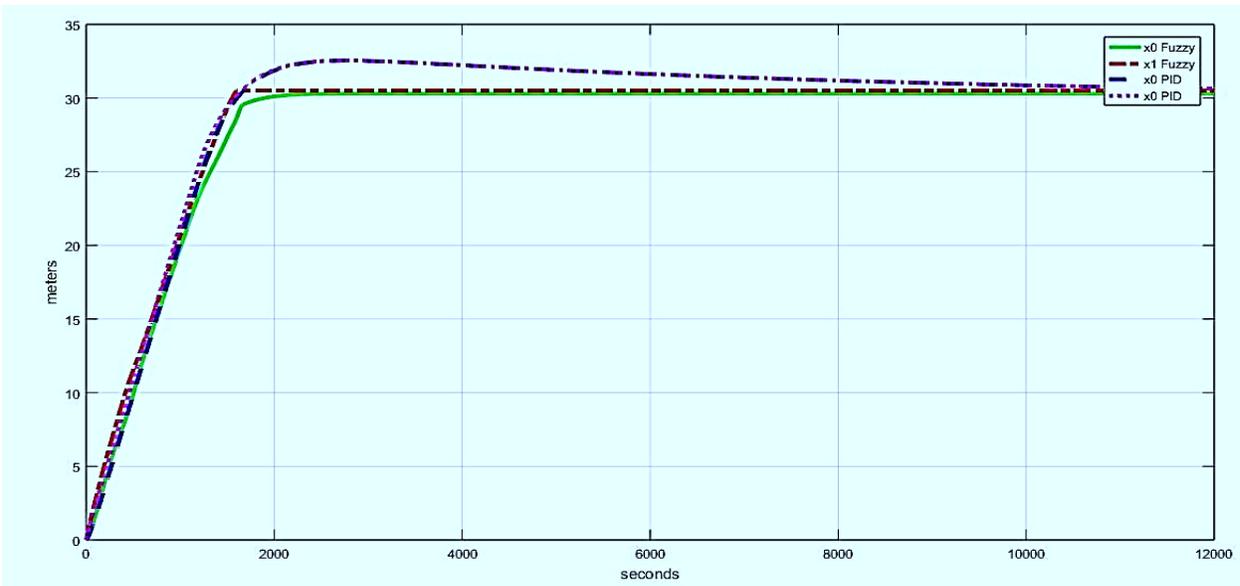

Figure 10 Comparison of Fuzzy with PID controller for the three pond system.

## 4.3 ROBUSTNESS ANALYSIS

In this section the robustness of proposed three pond model of ROR HPP regulated with FIS is analyzed and discussed. The system is simulated for its response against sinusoidal disturbances and flash flood surges.

### 4.3.1 Case III: Three pond system vs Single pond system for sinusoidal disturbance

While all the parameters of the traditional system are matched with our system so that there is a fair comparison. Since both of systems are expected to have similar times to reach the desired steady states, so they are compared against disturbances. An additive sinusoidal disturbance $d = a.sin(wt)$ is added to the system with the controllable inflow of water $U = [U_1, U_2]$. Both systems behave normally during the transient states as shown in Figure 11. The difference appears when the systems reaches the steady states, as the sinusoidal effect appears more pronounced in the traditional system while three pond system suppresses the effects of disturbance far better than traditional system. These results shows that the proposed model regulated with FIS is robust against the sinusoidal disturbances.

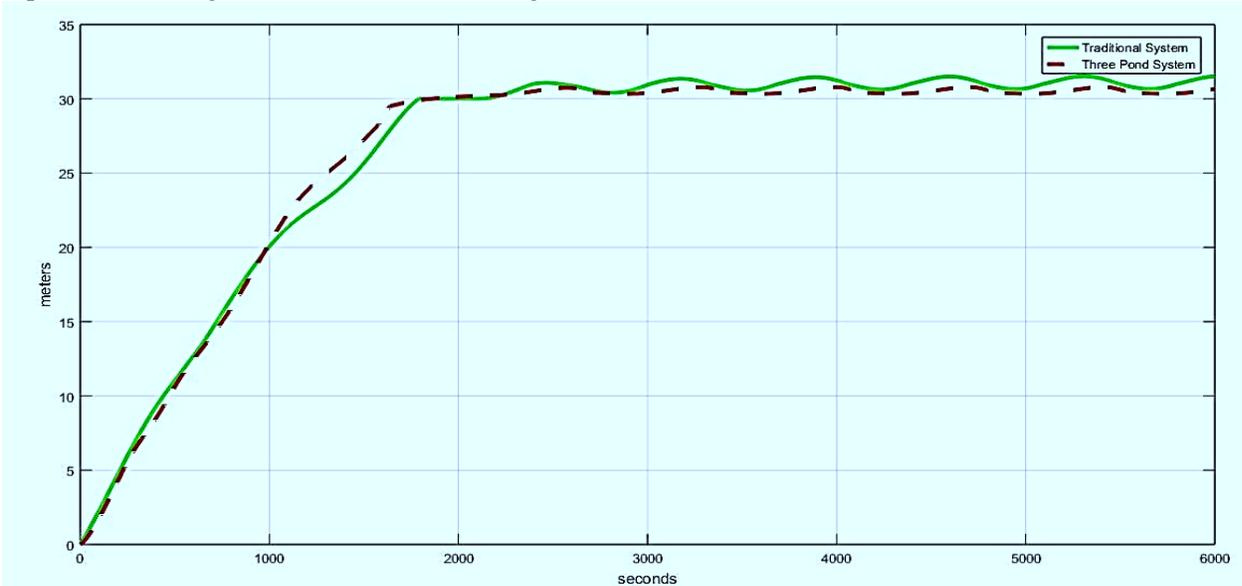

Figure 11 Comparison of proposed system with traditional system for a sinusoidal disturbance

### 4.3.2 Case IV: Three pond system vs. Single pond system for flash flood

The responses of the three pond and the traditional systems are tested against a flash flood and compared. As the systems reaches the steady state, a simulated flash flood or surge is added. The behaviors of the system is shown in Figure 12.

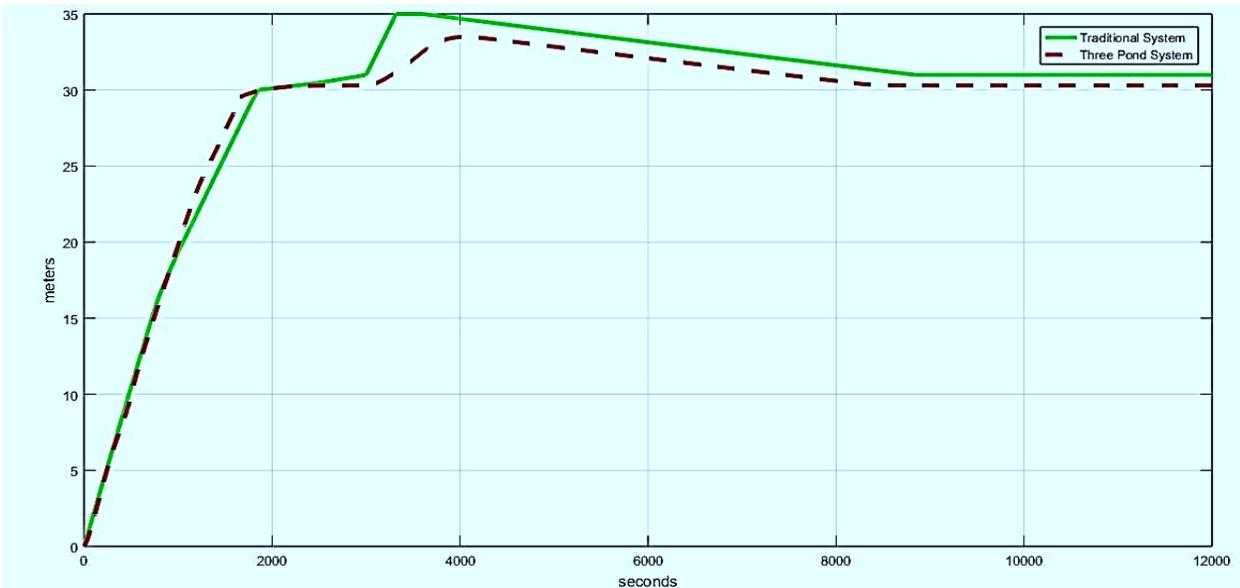

Figure 12 Comparison of proposed system with traditional system for a flash flood.

Three pond system (represented by red dotted line) deviates much lesser than the traditional system, moreover it also returns to steady state earlier. Furthermore, it is notable that the three pond system effectively damps and smoothens out the disturbance caused by the flash flood. These results shows that the proposed model regulated with FIS is robust against the flash flood surges.

## 5. CONCLUSION

This work presents a novel application of three pond model (inspired from three tank model) for run of river hydro power plants. The hydraulic model of three pond model is highly nonlinear in nature. Proposed scheme has applied PID controller and a Fuzzy inference system to a three pond model. The demand was to maintain the water level in ponds that results in sustainable energy generation. The results have shown that three pond model has achieved better results in terms of maintaining water level as compared to the traditional single pond model. Furthermore, it has also been observed that due to the highly non-linear nature of the problem Fuzzy Inference System (FIS) has achieved better results in maintaining desired water level. Finally the system has been tested against two different types of disturbance i.e. sinusoidal disturbance and flash flood disturbance. So the Fuzzy Inference System along with a three pond model has achieved robust results in terms of disturbance overshoots handling and settling time.

https://ieeexplore.ieee.org/abstract/document/1047196/

6. Belhadji L, Bacha S, … DR-I 2012-38th A, 2012 undefined. Experimental validation of direct power control of variable speed micro-hydropower plant. *ieeexplore.ieee.org*. Accessed September 8, 2020. https://ieeexplore.ieee.org/abstract/document/6388585/
7. Kim J, Gevorgian V, … YL-IT, 2019 undefined. Supercapacitor to Provide Ancillary Services With Control Coordination. *ieeexplore.ieee.org*. Accessed September 8, 2020. https://ieeexplore.ieee.org/abstract/document/8744562/
8. … IB-2017 6th I, 2017 undefined. Development of expert system for fault diagnosis of an 8-MW bulb turbine downstream irrigation hydro power plant. *ieeexplore.ieee.org*. Accessed September 8, 2020. https://ieeexplore.ieee.org/abstract/document/8003740/
9. Jiang J, Qiao Y, on ZL-2018 IC, 2018 undefined. The probabilistic production simulation for renewable energy power system considering the operation of cascade hydropower stations. *ieeexplore.ieee.org*. Accessed September 8, 2020. https://ieeexplore.ieee.org/abstract/document/8601991/
10. Molina M, on MP-2010 IIC, 2010 undefined. Improved power conditioning system of micro-hydro power plant for distributed generation applications. *ieeexplore.ieee.org*. Accessed September 8, 2020. https://ieeexplore.ieee.org/abstract/document/5472461/
11. Tessarolo A, Luise F, … PR-… C on C, 2011 undefined. Traditional hydropower plant revamping based on a variable-speed surface permanent-magnet high-torque-density generator. *ieeexplore.ieee.org*. Accessed September 8, 2020. https://ieeexplore.ieee.org/abstract/document/6036335/
12. Society GH-G-2011 IP and E, 2011 undefined. Predictive control for balancing wind generation variability using run-of-river power plants. *ieeexplore.ieee.org*. Accessed September 8, 2020. https://ieeexplore.ieee.org/abstract/document/6039197/
13. Borkowski D, Energy TW-IT on, 2013 undefined. Small hydropower plant with integrated turbine-generators working at variable speed. *ieeexplore.ieee.org*. Accessed September 8, 2020. https://ieeexplore.ieee.org/abstract/document/6479690/
14. Eslava Blanco HJ, Rojas Castellar LA, Herrera Sanabria MG. Intelligent integration based on optical telecommunications to optimize distributed generation. In: Institution of Engineering and Technology (IET); 2012:102-102. doi:10.1049/cp.2012.0751
15. Avram C, Mircescu D, … AA-2014 II, 2014 undefined. Fluid Stochastic Petri Nets based Modelling and simulation of Micro Hydro Power stations behaviour. *ieeexplore.ieee.org*. Accessed September 8, 2020. https://ieeexplore.ieee.org/abstract/document/6857889/
16. Xu H, Liu W, Wang L, … ML-… S on S, 2015 undefined. Optimal sizing of small hydro power plants in consideration of voltage control. *ieeexplore.ieee.org*. Accessed September 8, 2020. https://ieeexplore.ieee.org/abstract/document/7315201/
17. Wen X, Fan Q, Liu W, … CL-2018 2nd IA, 2018 undefined. Design and Application of Multiple Intermittent Energy Grid Integration Evaluation System. *ieeexplore.ieee.org*. Accessed September 8, 2020. https://ieeexplore.ieee.org/abstract/document/8469434/
18. Sabir M, Shah S, International UH-2017 3rd, 2017 undefined. Establishment of hydroelectric microgrids, need of the time to resolve energy shortage problems. *ieeexplore.ieee.org*. Accessed September 8, 2020. https://ieeexplore.ieee.org/abstract/document/8251794/
19. on DB-2015 16th ISC, 2015 undefined. Small hydropower plant as a supplier for the primary energy consumer. *ieeexplore.ieee.org*. Accessed September 8, 2020. https://ieeexplore.ieee.org/abstract/document/7161059/
20. Diniz A, … AS-2016 PS, 2016 undefined. An exact multi-plant hydro power production function for mid/long term hydrothermal coordination. *ieeexplore.ieee.org*. Accessed September 8, 2020. https://ieeexplore.ieee.org/abstract/document/7541013/
21. Chen Y, Liu F, Wei W, … SM-C journal of power, 2016 undefined. Robust unit commitment for large-scale wind generation and run-off-river hydropower. *ieeexplore.ieee.org*. Accessed September 8, 2020. https://ieeexplore.ieee.org/abstract/document/7785870/
16